\documentclass[ammsymb,prl,aps,twocolumn,superscriptaddress,showpacs]{revtex4}
\usepackage{graphicx}

\def \beq{\begin{equation}}
\def \eeq{\end{equation}}
\def \beqar{\begin{eqnarray}}
\def \eeqar{\end{eqnarray}}
\input{tcilatex}

\begin{document}

\title{The Evolutionary Minority Game with Local Coordination}
\author{E. Burgos}
\email{burgos@cnea.gov.ar, ceva@cnea.gov.ar, rperazzo@itba.edu.ar}
\affiliation{
Departamento de F{\'{\i}}sica, Comisi{\'o}n Nacional de Energ{\'\i }a At{\'o}%
mica, Avenida del Libertador 8250, 1429 Buenos Aires, Argentina}
\author{Horacio Ceva}
\affiliation{
Departamento de F{\'{\i}}sica, Comisi{\'o}n Nacional de Energ{\'\i }a At{\'o}%
mica, Avenida del Libertador 8250, 1429 Buenos Aires, Argentina}
\author{R.P.J. Perazzo}
\affiliation{
Dpto. de F\'{\i}sica, Facultad de Ciencias Exactas y Naturales,
Universidad de Buenos Aires and \\
Instituto Tecnol\'ogico de Buenos Aires, Departamento de Investigaci\'on y
Desarrollo, Avenida Eduardo Madero 399, Buenos Aires, Argentina}
\date{\today}

\begin{abstract}
We discuss a modification of the Evolutionary Minority Game (EMG) in which
agents are placed in the nodes of a regular or a random graph. A
neighborhood for each agent can thus be defined and a modification of the
usual relaxation dynamics can be made in which each agent updates her
decision depending upon her neighborhood. We report numerical results for
the topologies of a ring, a torus and a random graph changing the size of
the neighborhood. We find the surprising result that in the EMG a better
coordination (a lower frustration) can be achieved if agents base their
actions on local information disregarding the global trend in the self
segregation process.
\end{abstract}

\maketitle

\section*{INTRODUCTION}

There are a great number of situations in which a many agent system self
organizes by coordinating individual actions. Such coordination is usually
achieved by agents with partial information about the system, and in some
cases optimizing utility functions that conflict with each other. A similar
situation is found in many particles, physical systems. The word
``coordination'' used in a social or economic context is
then replaced by ``ordering''. Examples are the growth of a
crystalline structure or a transition leading to some specific magnetic
phase.

Interesting situations arise when the optimal configurations for different
individuals do collide with each other. This can be due to the nature of the
interactions between the particles as in a spin glass, or by boundary
conditions which prevent a global ordering or by the constitutive rules of a
system of multiple players that prevent that all agents can win. In these
cases it is said that the system displays some degree of \emph{frustration}.
An example of a frustrated multi agent system is given by the evolutionary
minority game (EMG) \cite{MGame} in which many players have to make a binary
choice and the winning option is the one made by the minority. The
similarities between some variants of the Minority Game and spin glasses
have been discussed in great detail in Ref. \cite{tano}.

A macroscopic signature of frustration is that the system can not
accommodate into a single, optimal state in which the energy is a minimum
but it relaxes instead to one of many, suboptimal, equivalent configurations
that are local minima in the energy landscape. In the random relaxation
dynamic that is used for the EMG each player continually modifies her choice
searching for a winning option. The final result is that the population is
segregated into two parties that take opposite actions. This partition is
not unique and also tends to reduce the frustration as much as possible by
minimizing the number of losers.

The relaxation process is usually assimilated to the search of a solution of
a combinatorial optimization problem in which it is possible a strategy of
``divide and conquer'' \cite{libro}, i.e. circumstances in
which one can attempt to divide the system into parts and search for
separate optima in each part. Frustration arises when such local solutions
can not be reassembled into a global optimum also fulfilling the boundary
conditions.

A relevant example of the study of the global outcome of a local
coordination strategy (i.e. involving only a fraction of the system) is
Schelling's segregation model \cite{schelling}. Agents of two kinds are
placed in a square grid. The system relaxes to equilibrium allowing any two
agents of different kinds to exchange places if they are surrounded by, say,
a majority of agents of the opposite kind. In the present paper we discuss a
relaxation dynamics for the EMG in which we impose a local coordination
strategy. We borrow the picture of Schelling's models and place the players
in a lattice. It is then possible to associate a neighborhood to each player
and thus implement a local coordination strategy letting each player to
adjust her decision to the situation in her neighborhood. We call this model
the Local Evolutionary Minority Game (LEMG).

A previous work in this direction is \cite{suiza} in which players are also
located on the nodes of a grid but are endowed with (two) strategies that
are selected on the basis of their successful use. This work further imposes
that both strategies have to be anticorrelated, i.e. they tend to produce
opposite actions with the same input. We stress the \emph{evolutionary}
nature of the present model: players bear no memory of past actions and do
not have any strategy in the sense of \cite{suiza}, to guide their actions.
In spite of this difference the effects of local coordination produce a
similar ordered pattern. We pay special attention to the effects of such a
local coordination in the optimization process.

\section{The rules of the LEMG}

We first consider the traditional EMG. This involves $N$ players that make
one binary decision (0 or 1). Each player has a probability $%
p_{i};i=1,2,\cdots ,N$ of choosing, say, 0. Each player receives one point
if her decision places her in the minority, and loses a point otherwise.
When her account of points falls below 0, she changes $p_{i}\rightarrow
p_{i}^{\prime }$ with $p_{i}^{\prime }\in \lbrack p_{i}-\Delta
p,p_{i}+\Delta p]$, at random, and $\Delta p\ll 1$. Reflective boundary
conditions are imposed at $p_{i}=0,1$. All agents are assumed to update the
corresponding $p_{i}$'s synchronically. It is customary to display the
self-organization of the system through the probability density function $%
P(p)$ obtained in a statistical ensemble of systems that are allowed to
relax to equilibrium. The value of $P(p)dp$ is the fraction of the
population having a probability between $p$ and $p+dp$ of choosing, say, 0.
When the probabilistic relaxation is used, the asymptotic function $P(p)$
has a U-shape with two symmetric peaks at $p\simeq 0$ and $p\simeq 1$ thus
indicating that the $N$ agents have segregated into two parties making
opposite decisions. The relaxation process corresponds to the minimization
of an ``energy'' function \cite{Thermal}
given by the standard deviation $\sigma $ defined by:
\begin{equation}
\sigma ^{2}=\sum_{A}\mathcal{P}(A)(A-N/2)^{2}  \label{costo}
\end{equation}%
where $\mathcal{P}(A)$ is the probability distribution of parties of $A$
agents that have chosen 0. The value of $\sigma ^{2}$ depends upon the
properties of $P(p)$. In \cite{Thermal} it is proven that:
\begin{equation}
\mathcal{E}\equiv \frac{\sigma ^{2}}{N}=N(\bar{p}-1/2)^{2}+(\overline{p}-%
\overline{p^{2}})  \label{costo2}
\end{equation}%
where $\overline{p^{s}}=\int p^{s}P(p)dp$. At equilibrium the linear
dependence of $\mathcal{E}$ on N disappears, and $\sigma ^{2}$ turns out to
be an extensive magnitude proportional to $N$. A minimization of $\mathcal{E}
$ is equivalent to find a distribution $P(p)$ with the smallest possible
number of losers. In fact $\sigma ^{2}$ is related to the number of losers
because
\begin{equation}
\sigma ^{2}=<(A-N/2)^{2}>=\frac{<(w-\ell )^{2}>}{4}=\frac{<(N-2\ell )^{2}>}{4%
} \label{ec.3}
\end{equation}%
where $w$ ($\ell $) is the number of winners (losers) and $<\cdots >$
represents an ensemble average.

If one assumes na{\"{\i}}vely $P(p)=\delta (p-1/2)$ corresponding to a
symmetric random walk (and thus eliminating the term $O(N)$ in Eq.[\ref%
{costo2}]) one gets $\mathcal{E}=1/4,$ while $P(p)=$ constant yields $%
\mathcal{E}=1/6$. A better result is obtained with the usual random
relaxation dynamics for the EMG. This yields \cite{Thermal} $\mathcal{E}%
\simeq 1/8$. Energy and frustration remain linked to each other. For the EMG
we can define frustration as $\mathcal{F}=\ell /N$; which fulfills $0\leq
\mathcal{F}\leq 1$. This definition may also be used for any system
involving a game with multiple players. The value $\mathcal{F}=0$
corresponds to a situation such as the  ``majority game'' in which a player 
is a winner if her decision is the
same as the majority. This leads to situations that can be assimilated to a
ferromagnetic phase (all the players (spins) have chosen the same option
(orientation)). In the EMG there are less winners than losers, and therefore
$1/2<\mathcal{F}_{EMG}\leq 1$. The lowest possible frustration for the EMG
is reached when the $N$ (odd) agents are coordinated to produce the largest
possible minority, i.e. $(N-1)/2$. Thus the lowest possible frustration for
a finite minority game is $\mathcal{F}^{\ast }=(1+1/N)/2$.


We now turn to the LEMG in which the $i$-th player makes her decision
depending upon the situation in her neighborhood $\mathcal{N}_i$. In order
to define the neighborhoods we assume three possible spatial orderings. Two
of them correspond respectively to a one-dimensional (1D) or a square two-
dimensional (2D) regular array with periodic boundary conditions (i.e.
respectively a ring and a torus). In the third arrangement the agents are
placed in the nodes of a random undirected graph with a fixed number of
neighbors for each agent so that a reciprocity relationship is automatically
fulfilled (if node $i$ is taken to be linked to node $j$, the reciprocal is
also true). All neighborhoods are assumed to have the same (odd) number $n$
of agents (we consider that $i \in \mathcal{N}_i$).

The rules of the LEMG are the same as for the EMG except for the important 
difference that an agent wins or loses points depending whether she is, or 
she is not, in the minority \emph{of her own neighborhood}. No attention is 
paid to the agents that do not belong to $\mathcal{N}_i$. The LEMG 
coincides with the usual EMG when $\mathcal{N}_i$ coincides with the 
complete $N$ -agent system. In the regular orderings the neighborhoods are 
respectively a segment or a square with an odd number of agents. The only 
agent that updates her $p_i$ is located at the center of the square or 
segment. Notice that an agent may be in the minority (a winner) in her 
neighborhood and in the majority (a loser) when the entire system is 
considered, and vice versa.

Let us consider the simple example of an infinite linear chain of agents and
a neighborhood with $n=3 $. We define $R_i$ to be the probability that the $%
i $-th agent belongs to the minority of $\mathcal{N}_i$. We can thus write:
\begin{equation}
R_i= (1-p_{i-1})p_i(1-p_{i+1})+p_{i-1}(1-p_i)p_{i+1}  \label{probentorno}
\end{equation}
where $-\infty \leq i \leq \infty $. The probability that all agents are
winners is $R=\prod_i R_i $. Obviously $R=1$ if and only if $R_i=1,\forall i$%
. This is possible only if $p_ i=1$ and $p_{i\pm1}=0$. This corresponds to a
pattern in which 0's and 1's alternate with a period of 2. Larger
neighborhoods give also rise to periodic solutions with larger (even)
periods. Any finite ring with an odd number of agents is frustrated because
such periodicity can not fit along the chain.

Equation [\ref{probentorno}] can be used to construct a (deterministic)
relaxation dynamics to adjust the $p_{i}$'s climbing along the gradient of $%
R_{i}$. We thus assume $p_{i}(t+1)=p_{i}(t)+\Delta p_{i}$ and set
\begin{equation}
\Delta p_{i}=\eta \partial R_{i}/\partial p_{i}=\eta (1-p_{i-1}-p_{i+1})
\end{equation}%
with $1\gg \eta >0$. A stationary ($\Delta p_{i}=0\ \forall i$) solution for
this dynamics is $p_{i}=1/2\ ;\ \forall i$. This solution is unstable
because any random perturbation of any $p_{i}$ leads to a situation in which
$\Delta p_{i}\not=0\ \forall i$ \cite{foot2}. This dynamics stabilizes a
pattern of 0's and 1's that alternate with each other. In fact if $p_{i+1}$
and $p_{i-1}$ are both greater (smaller) than 1/2, then $\Delta p_{i}<0(>0)$
thus forcing $p_{i}<1/2(>1/2)$. This relaxation dynamics is therefore
expected to lead to distributions $P(p)$ that vanish at $p=1/2$.

\section{RESULTS}

\subsection*{Self-segregation}

\begin{figure}[tbp]
\includegraphics[width=9cm,clip]{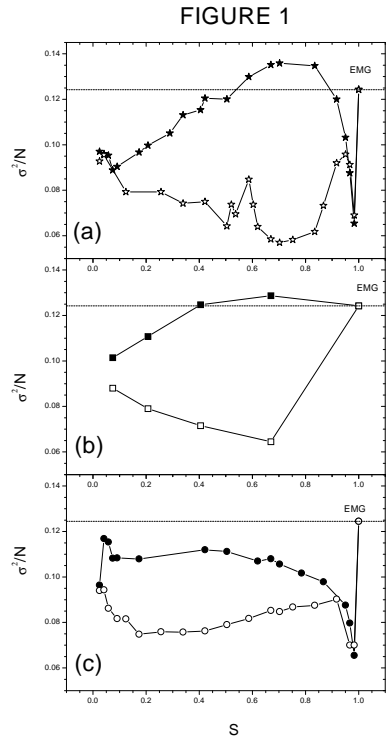}
\caption{ $\protect\sigma^2/N$ as as a function of the size parameter $S$
for different topologies of the $N=121$ players system. Panel a) corresponds
to a ring, panel b) to a square torus and panel c) to a random graph. Lines
are drawn to guide the eye. Empty and filled symbols correspond respectively
to results obtained with and without annealing (see the text). Data was
obtained from 200 independent histories, of $5 \times 10^5$ time steps each,
$\Delta p=0.1$ and by averaging over the last 2000 time steps of all the
histories. The annealing protocol consists in resetting all accounts to
zero every 500 iterations. This is repeated 800 times. The fluctuations
observed in the lower curve of panel a) subsist in simulations with a much
richer statistics (see the text).}
\label{energia}
\end{figure}

In Fig.\ref{energia} we show the results of $\mathcal{E}=\sigma ^{2}/N$ as a
function of the size parameter $S$ defined as the ratio $S=n/N$ obtained in
several numerical experiments. The value for $S=1$ corresponds to $\mathcal{E%
}_{EMG}$ obtained for the EMG. In this section we only discuss results for $%
N=121$. We have considered the topologies of a ring, a ``square'' torus 
(with $N=11\times 11$), and of random
graphs. In all the cases considered, the values of $\mathcal{E}_{S}$ with $%
S\ll 1$ fulfill $\mathcal{E}_{S}<\mathcal{E}_{EMG}$. This feature is
stressed in Fig.\ref{energia} with an horizontal line drawn at the value of $%
\mathcal{E}_{EMG}$. The value of $\mathcal{E}_{S}$ for regular 1D and 2D
lattices grows with $S$ and for $S\simeq .5$ becomes even larger that $%
\mathcal{E}_{EMG}$.

\begin{figure}[tbp]
\includegraphics[width=9cm,clip]{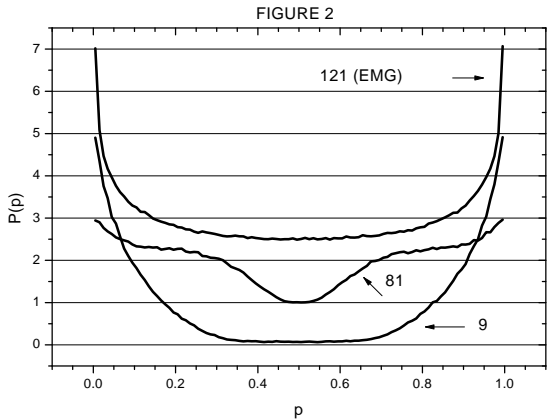}
\caption{ Examples of the density distribution $P(p)$ for the linear chain
and different size parameters ($S=9/121$, $S=81/121$ and $S=1$ (EMG)). The
second and third curves are offset by one and two units, respectively.
Notice that the first two curves (almost) drop to zero around $p\simeq 0.5$}
\label{P(p)}
\end{figure}

The corresponding density distributions $P(p)$ are shown in Fig.\ref{P(p)}
for several values of $S$. These are compared with the distribution $%
P_{EMG}(p).$ We observe that $P_{S}(p)$ with $S\ll 1$ are always symmetric
and U-shaped as $P_{EMG}(p)$ but they differ from this in the fact that they
vanish around $p=1/2$. This agrees with the discussion given above for the
linear chain. As we shortly discuss, this turns out to be a highly relevant
and general feature of the LEMG. Such distributions are a better
approximation to an ideal distribution
\begin{equation}
P(p)=\frac{N-1}{2N}[\delta (p)+\delta (p-1)]+\frac{1}{N}\delta (p-1/2)
\end{equation}%
that yields the optimal value of $\mathcal{E}_{EMG}^{\ast }=1/4N$ (and $%
\mathcal{F}_{EMG}^{\ast }=(1+1/N)/2$). A noticeable dip is produced for $%
S=(N-2)/N$. This can be understood in the following way. Assume that a
symmetric distribution $P(p)$ has already developed and two agents are
removed in order that the $i-$th player can check her decision in her
neighborhood of $N-2$ agents. If the two agents that have been removed have $%
p>1/2$ ($p<1/2$) the $i-$th agent has the single winning option of choosing $%
p_{i}\simeq 1$ ($p_{i}\simeq 0$). In the other cases (one player with $p>1/2$
and the other with $p<1/2$) her choices of approaching 0 or 1 have equal
probability. The net result once all players have updated her respective $p$%
's in the same fashion is to force the resultant $P(p)$ to drop at p=1/2 and
grow at $p=0,1$ as discussed for the linear chain. This argument can be
extended for neighborhoods of other sizes.

\subsection*{The Optimization Problem}

\begin{figure}[tbp]
\includegraphics[width=9cm,clip]{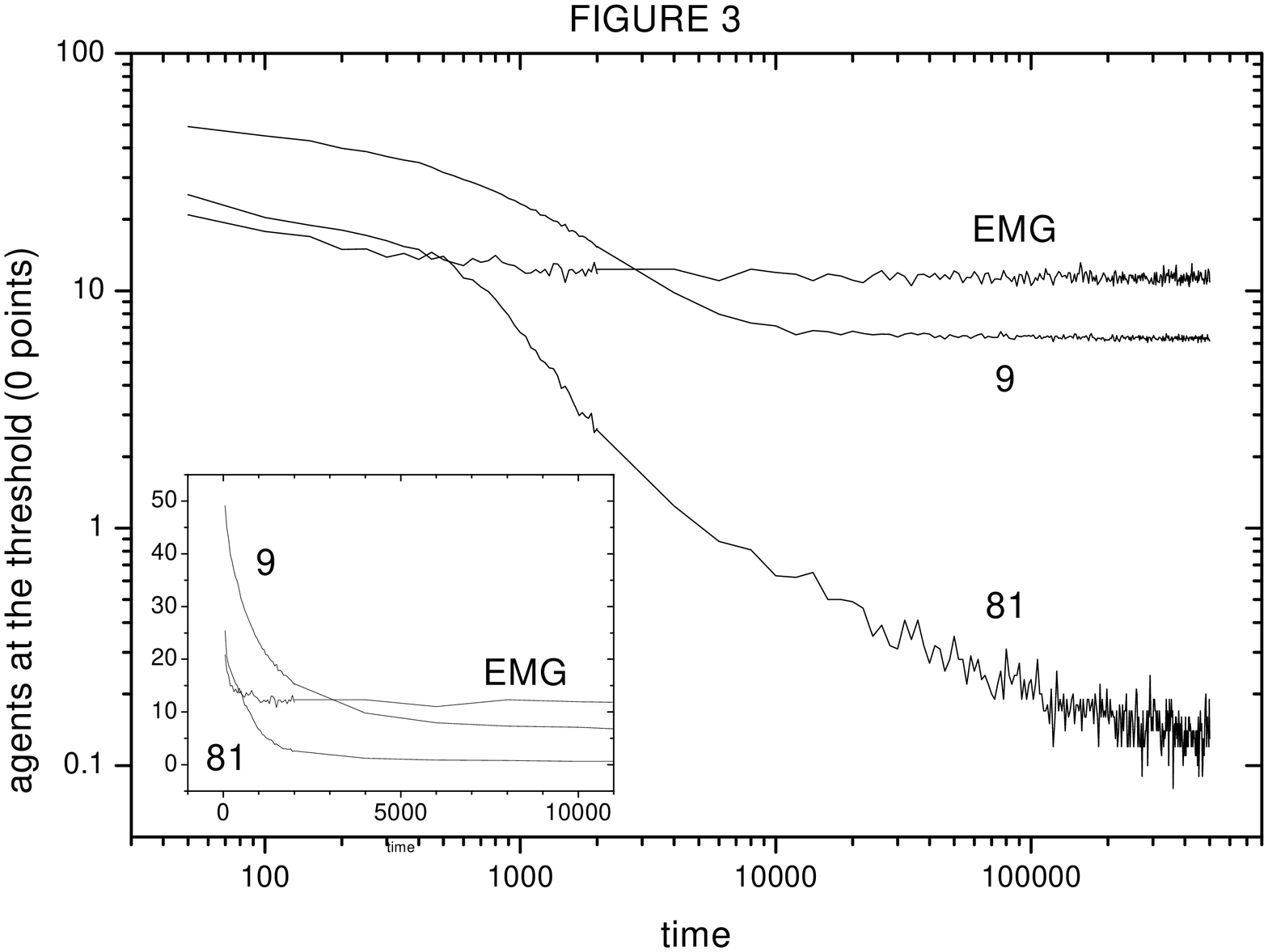}
\caption{Number of players with zero points as a function of the iteration
number (time) for the same values of the size parameter as in Fig.\protect\ref{P(p)}
. In the inset we display in a linear scale the same data up to 10,000 time
steps, to put in evidence the presence of a fast and a slow dynamics}
\label{descontentos}
\end{figure}

The shape of $P_{S}(p)$ changes with $S$. For intermediate values (for
instance $S=81/121$) this distribution has radically changed from the shape of a U
to a two wing profile with secondary maxima at both sides of $p=$1/2 still
keeping the fact that $P_{S}(1/2)=0$. This is associated to an increase in
the number of ``local winners'' . In fact
after some time there are left almost no players that need to update their $p_{i}$'s 
(see Fig.\ref{descontentos}) while for $S \simeq 0$ or $S\simeq 1$
the relaxation process reaches a dynamical equilibrium in which few players
continuously update their $p_{i}$'s. 

For such intermediate values of $S$ it is found that a minority is clearly 
defined in most neighborhoods and the corresponding agents are unambiguously 
induced to take one winning option. They therefore continue to accumulate 
points and cease to change their $p_{i}$'s preventing the system to reach 
a more efficient self-segregation. 

A situation like this has been extensively discussed in 
Ref.\cite{Quenching}. In these circumstances the relaxation process ceases 
to be effective to lower the energy and the system freezes in a 
configuration that is far from a better \emph{local} optimum. Although the frozen 
microstates do depend upon initial conditions, the asymptotic density 
$P(p)$ is independent both because this is a density function that is 
associated with a macrostate of the system and because possible random 
variations are averaged out by repeating the relaxation process for a large 
ensemble of systems. Moreover, succesive runs to obtain ${\cal E}$ changing 
the random initial configuration, yield essentially the same result. This 
is indeed to be expected because the energy is defined as the average of a 
statistical fluctuation, as given in Eq.[\ref{ec.3}].

The procedure to regain the true optimal self-segregation pattern, is to force 
the relaxation procedure by periodically removing the points that have been 
accumulated by every player, resetting their accounts to 0. This procedure
changes \emph{only} the situations in which the system is frozen but leaves
unchanged situations in which this does not happen such as for instance
for $S \simeq 0$ or $S \simeq 1$. 

These  ``annealing'' episodes melt the 
system thus making it possible to reach the best local configurations. We 
have performed an annealed relaxation (with a fixed annealing protocol) for 
all three topologies. The results are displayed with open symbols in all 
three panels of Fig.\ref{energia}. The fluctuations in the lower curve of 
panel a) for $S\simeq 0.6$ disappear for 
$N\geq 500$ thus indicating that it is a finite size artifact of the model.

A remarkable result displayed in Fig.\ref{energia} is that the composition
of local optima always yields a better coordination than the one
obtained within the framework of the EMG in which all agents are involved in
the same relaxation process. 

There are actually two situations to consider. One in which $S \simeq 0$ or $S 
\simeq 1$, and the other where $S$ is within these two extreme values. In the 
first case the LEMG always yields remarkably lower values of ${\cal E}$. This is 
indeed a general result, because holds true no matter the topology and the size 
of the system in which the players are located. In these cases no annealing 
is required because the system never gets quenched.  

Outside the neighbourhood of  $S \simeq 0$ or $S \simeq 1$, the system gets 
quenched, and for regular topologies a poorer value of ${\cal E} $ 
(\emph{i.e.} greater than  ${\cal E}_{EMG}$) is obtained. However, for such 
values a better (lower) value of ${\cal E}$ can always be obtained if
the annealing procedure is used to force the relaxation 
process beyond the frozen states. This actually means that a better 
coordination, i.e. a lower frustration, is achieved whenever the system is 
found in a configuration that corresponds to a composition of good local 
optima, and this is true in the whole range of values of $S$. If good local 
optima are guaranteed one finds values of ${\cal E}_{S}$ that are 
significantly lower that ${\cal E}_{EMG}$. A typical value of ${\cal 
E}_{S}\simeq$ 1/16 is obtained in this way that is half the
value ${\cal E}_{EMG}\simeq$ 1/8.

Except for finite size effects, all the results presented in this subsection
do not differ from those obtained for many other values of $N$ that we have
investigated.

\subsection*{The 2D Case}

A much richer situation is found for the case in which players are located
on a grid with the topology of 2D torus. An example is shown in Fig.\ref%
{bidimensional} in which $N=31\times 31$. The values of $p$ are associated
to shades of gray. Frustration can be perceived in the fact that there is
not a single global ordering of black ( $p\simeq 0$) and white ($p\simeq 1$)
stripes for the whole array. These are instead grouped in domains with
different orientations or with the same orientation but shifted with respect
to each other. The relaxation process is fast in an initial stage and slows
down once the domains have fully developed. The domain walls are a source of
frustration. In fact, when such stage has been reached all the agents that
have 0 points and continue to update their $p_{i}$'s are located in the
domain walls giving rise to a slow dynamics in which walls move enlarging or
shrinking domains. The whole picture resembles a crystallization process;
for a value of $S$ grater than a critical threshold, all domains collapse
into a single pattern of stripes. Frustration shows up as an indented
(fuzzy) border in some of the stripes. These results will be discussed in
detail elsewhere.

\begin{figure}[tbp]
\includegraphics[width=9cm,clip]{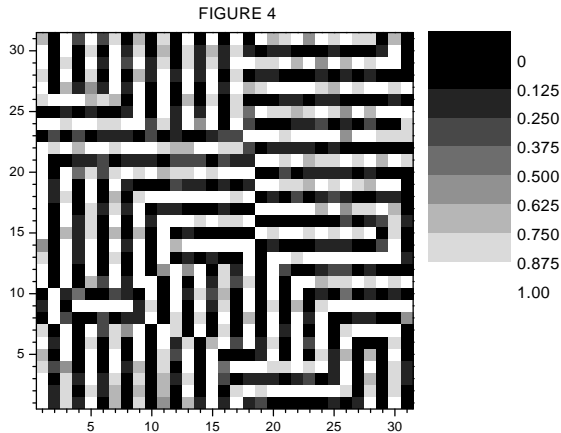}
\caption{An example of the LEMG for the topology of the torus.  Domains in
the map of probabilities for a $3 \times 3$ neighborhood.  Each pixel
represents a player; the corresponding $p_i$ are shown as shades of gray.}
\label{bidimensional}
\end{figure}

\section{Conclusions}

We have studied the organization pattern achieved by many agents playing an
EMG with local coordination. We find important differences between the
coordination achieved when the whole ensemble of agents participates in the
same relaxation process or when local coordination is imposed. 

According to the present results, the LEMG is an example in which a better
coordination or, what is the same, a lower frustration, can be achieved
provided that the self organization is ruled by a local process, i.e. if
agents govern their actions paying no attention to global trends of the
system but rather to her immediate neighborhood. In the LEMG, this statement
holds true even in the case in which very few agents are removed ($S\simeq
(N-2)/N$ ) from the whole ensemble. The fact that a more efficient
coordination is achieved by \emph{ignoring} what happens to the total
ensemble of players is expected to be a feature of a special kind of
multi-agent systems. Other coordination problems may not behave in this
fashion, thus allowing a classification of coordination games into classes
linked to the type of optimization problem that is being ``solved''
by the ensemble of agents. Further investigation
should be devoted to classify multi agent games into those that fulfil this
property and those that can not be optimized by braking them into pieces.

It is found that the LEMG displays some of the features that are typical of
antiferromagnetic systems including the emergence of domains, frustration,
fast and slow dynamics, etc., while also keeping the essence of multiagent
models, as applied to social or economic organization. It therefore sheds
light on the connection between those two bodies of knowledge. E.B. has been
partially supported by CONICET of Argentina, PICT-PMT0051.


\begin{thebibliography}{9}
\bibitem{foot2} There are other stationary solutions, such as a saw-tooth
profile repeating the pattern $p_{i \pm1} = 1/2 \mp \epsilon$ and $p_{i} =
1/2$. These solutions are also unstable
\end{thebibliography}
\end{document}